\documentclass[12pt]{article}
\usepackage[totalwidth=450pt, totalheight=590pt, centering]{geometry}

\usepackage{amsmath,amssymb,mathrsfs}
\usepackage{bbm} 

\usepackage[dvips]{graphicx}

\usepackage[dvips,bookmarks=false]{hyperref} 
\hypersetup{pdfstartview=FitH,pdfhighlight=/O,colorlinks=false}

\newcommand{\mc}[1]{\mathcal{#1}}

\newcommand{\mbb}[1]{\mathbb{#1}}
\newcommand{\mbf}[1]{\mathbf{#1}}

\newcommand{\Tr}{{\rm Tr}}

\newcommand{\mdots}{,.\,.\,,}

\title{\Large{\textbf{Coherent spin-networks}}}

\author{Eugenio Bianchi{\it ${\,}^{a}$}, Elena Magliaro{\it ${\,}^{a}$}, Claudio Perini{\it ${\,}^{ab}$}\\[.35em]
\small{\textit{${}^a$Centre de Physique Th\'eorique de Luminy}}\footnote{Unit\'e mixte de recherche (UMR 6207) du CNRS et des Universit\'es
de Provence (Aix-Marseille I), de la M\'editerran\'ee (Aix-Marseille II) et du Sud (Toulon-Var); laboratoire affili\'e \`a la FRUMAM (FR 2291).}\small{\textit{, case 907, F-13288 Marseille, EU}}\\
\small{\textit{${}^b$Dipartimento di Matematica, Universit\`a degli Studi Roma Tre, I-00146 Roma, EU}}
}

\date{June 2, 2010}

\begin{document}

\maketitle

\begin{abstract}
In this paper we discuss a proposal of coherent states for Loop Quantum Gravity. These states are labeled by a point in the phase space of General Relativity as captured by a spin-network graph. They are defined as the gauge invariant projection of a product over links of Hall's heat-kernels for the cotangent bundle of $SU(2)$. The labels of the state are written in terms of two unit-vectors, a spin and an angle for each link of the graph. The heat-kernel time is chosen to be a function of the spin. These labels are the ones used in the Spin Foam setting and admit a clear geometric interpretation. Moreover, the set of labels per link can be written as an element of $SL(2,\mbb{C})$. These states coincide with Thiemann's coherent states with the area operator as complexifier. We study the properties of semiclassicality of these states and show that, for large spins, they reproduce a superposition over spins of spin-networks with nodes labeled by Livine-Speziale coherent intertwiners. Moreover, the weight associated to spins on links turns out to be given by a Gaussian times a phase as originally proposed by Rovelli.
\end{abstract}


\section{Introduction}
In Loop Quantum Gravity (LQG) \cite{LQG,Smolin:2004sx,Ashtekar:2004eh,Thiemann:2007zz}, the recent convergence of the canonical and the covariant (Spin Foam) formulation \cite{Engle:2007wy,Freidel:2007py,Kaminski:2009fm} is shedding a new light on the identification of the classical regime of the theory. A key ingredient in this analysis is \emph{semiclassical states}, that is states peaked on a prescribed intrinsic and extrinsic geometry of space. 

In the recent graviton propagator calculations \cite{Rovelli:2005yj,Bianchi:2006uf,Livine:2006it,Alesci:2007tx,Alesci:2007tg,Alesci:2008ff,Bianchi:2009ri}, semiclassical states associated to a spin-network graph $\Gamma$ are considered. In particular, the states used in \cite{Bianchi:2009ri} are labeled by a spin $j^0_e$ and an angle $\xi_e$ per link $e$ of the graph, and for each node a set of unit vectors $\vec{n}$, one for each link at that node. Such variables are suggested by the simplicial interpretation of these states: the graph $\Gamma$ is in fact assumed to be dual to a simplicial decomposition of the spatial manifold, the vectors $\vec{n}$ are associated to unit-normals to faces of tetrahedra, and the spin $j^0_e$ is the average of the area of a face. Moreover, the simplicial extrinsic curvature is an angle associated to faces shared by tetrahedra and is identified with the label $\xi_e$. Therefore, these states are labeled by an intrinsic and extrinsic simplicial $3$-geometry.

More in detail, the semiclassical states used in \cite{Bianchi:2009ri} are obtained via a superposition over spins of spin-networks having nodes labeled by Livine-Speziale coherent intertwiners \cite{Livine:2007vk,Conrady:2009px,Freidel:2009nu}. The coefficients $c_j$ of the superposition over spins are given by a Gaussian times a phase as originally proposed by Rovelli in \cite{Rovelli:2005yj}
\begin{equation}
c_j(j_0,\xi)=\exp\big(-\frac{(j-j_0)^2}{2 \sigma_0}\big)\,\exp(-i \xi j)\;.
\label{eq:cj}
\end{equation}
Such proposal is motivated by the need of having a state peaked both on the area \emph{and} on the extrinsic angle. The dispersion is chosen to be given by $\sigma_0\approx (j_0)^k$ (with $0<k<2$) so that, in the large $j_0$ limit, both variables have vanishing relative dispersions (as explained in \cite{Bianchi:2006uf}). Moreover, a recent result of Freidel and Speziale strengthens the status of these classical labels \cite{SpezialeBeijing,newSpezialeFreidel}: they show that the phase space associated to a graph in LQG can actually be described in terms of the labels $(j^0_e, \xi_e,\vec{n}_e,\vec{n}'_e)$ associated to links of the graph. \\

While the states discussed above have good semiclassical properties and a clear geometrical interpretation, finding a better top-down derivation of the coefficients (\ref{eq:cj}) is strongly desirable. The reason is not just aesthetic. A derivation generally comes together with an understanding of its origins and with new mathematical tools that allow to simplify calculations. This is one of the objectives of this paper. \\

On the other hand, within the canonical framework, Thiemann and collaborators have strongly advocated the use of complexifier coherent states \cite{Thiemann:2000bw,Thiemann:2000ca,Thiemann:2000bx,Thiemann:2000by,Sahlmann:2001nv,Thiemann:2002vj,Bahr:2007xa,Bahr:2007xn,Flori:2008nw,Flori:2009rw}. Such states are labeled by a graph $\Gamma$ and by an assignment of a $SL(2,\mbb{C})$ group element to each of its links. The state is obtained from the gauge-invariant projection of a product over links of modified\footnote{In particular, the modified heat-kernels reduce to ordinary Hall's heat-kernels for the complexification of $SU(2)$ when the complexifier is chosen to be the area operator.} heat-kernels for the complexification of $SU(2)$. Their peakedness properties and geometric interpretation within the canonical theory have been studied in detail \cite{Thiemann:2000ca,Bahr:2007xn}. However the interpretation of the $SL(2,\mbb{C})$ labels in terms of discrete geometries and the relation with semiclassical states used in Spin Foams has largely remained unexplored. Exploring these aspects is the other objective of this paper.\\

Surprisingly, the two goals discussed above turn out to be strictly related. In this paper we present a proposal of \emph{coherent spin-network states}: the proposal is to consider the gauge invariant projection of a product over links of Hall's heat-kernels for the cotangent bundle of $SU(2)$ \cite{Hall1994,Ashtekar:1994nx}. The labels of the state are the ones used in Spin Foams: two normals, a spin and an angle for each link of the graph. This set of labels can be written as an element of $SL(2,\mbb{C})$ per link of the graph. Therefore, these states coincide with Thiemann's coherent states with the area operator chosen as complexifier, the $SL(2,\mbb{C})$ labels written in terms of the phase space variables $(j^0_e, \xi_e,\vec{n}_e,\vec{n}'_e)$ and the heat-kernel time given as a function of $j^0_e$.

We show that, for large $j^0_e$, coherent spin-networks reduce to the semiclassical states used in the spin-foam framework. In particular we find that they reproduce a superposition over spins of spin-networks with nodes labeled by Livine-Speziale coherent intertwiners and coefficients $c_j$ given by a Gaussian times a phase as originally proposed by Rovelli. This provides a clear interpretation of the geometry these states are peaked on.

\section{Coherent spin-network states}
The Hilbert space of LQG decomposes into sectors $\mc{K}_\Gamma=L^2(SU(2)^L/SU(2)^N,d\mu^L)$ associated to an embedded graph $\Gamma$ having $L$ links and $N$ nodes. States $\Psi(h_1\mdots h_L)$ in $\mc{K}_\Gamma$ capture a finite number of degrees of freedom of General Relativity: the ones associated to the classical phase space $\big(T^*SU(2)\big)^L$ of holonomies of the Ashtekar-Barbero connection along links of the graph and fluxes through surfaces dual to links of the graph. Here we consider states belonging to $\mc{K}_\Gamma$ and labeled by a point in phase space. Notice that the cotangent bundle $T^*SU(2)$ is diffeomorphic to the group $SL(2,\mbb{C})$. \footnote{In fact, since $SU(2)$ is a Lie group, its tangent bundle is trivial: $T^*SU(2)\simeq SU(2)\times su(2)^*\simeq SU(2)\times su(2)$; then observe that every element in $SL(2,\mbb{C})$ is of the form $x\,\exp(i y)$ with $x\in SU(2)$ and $y\in su(2)$. Moreover, the complex structure of $SL(2,\mbb{C})$ and the symplectic structure of $T^*SU(2)$ fit together so as to form a K\"ahler manifold.} This fact is largely exploited in the following: the states we consider are in fact labeled by an element of $SL(2,\mbb{C})$ per link of the graph.\\

Let us consider the heat kernel $K_t(h,h_0)$ on $SU(2)$. It has the following Peter-Weyl expansion
\begin{equation}
K_t(h,h_0)=\sum_j (2j+1) e^{-j(j+1)t}\, \chi^{(j)}(h\, h_0^{-1})\;.
\end{equation}
It is easy to show that, as a function of $h$, it is peaked on the conjugacy class of $h_0$. Moreover, when seen as a LQG state associated to a loop $\gamma$, $\Psi_\gamma(h)=K_t(h,h_0)$, it is peaked on small areas (that is small $j$). There is a rather simple variant of $K_t(h,h_0)$ that allows to peak on a prescribed spin $j_0$: it is given by the complexified heat kernel, i.e. by $K_t(h,H_0)$ with the group element $H_0$ belonging to $SL(2,\mbb{C})$. This is the unique analytic continuation of the $SU(2)$ heat kernel. These objects are the building blocks of coherent spin-networks. 

To simplify the notation, in the following we assume that $\Gamma$ is a complete graph so that, if $a,b,.\,.=1\mdots N$ label nodes of the graph, then links are labeled by couples $ab$. For instance, the holonomy associated to an oriented link is $h_{ab}$ and its inverse is $h_{ba}$. The generalization to arbitrary graph is immediate.

Coherent spin-networks are defined as follows: we consider the gauge-invariant projection of a product over links of heat kernels,
\begin{equation}
\Psi_{\Gamma,H_{ab}}(h_{ab})=\int\big(\prod_a dg_a\big)\;\prod_{ab} K_{t_{ab}}(h_{ab},\,g_a\, H_{ab}\, g_b^{-1}).
\label{eq:proposal}
\end{equation}
These states were first considered in \cite{Thiemann:2000bw}. The positive numbers $t_{ab}$ can be fixed in terms of the labels $H_{ab}$ as explained later on. Now, notice that every element $H_{ab}$ of $SL(2,\mbb{C})$ can be written in terms of two unit-vectors in $\mbb{R}^3$, a positive real number and an angle, that is: exactly the labels used in the semiclassical states adopted in spin foams \cite{Bianchi:2009ri}, the ones that in a simplicial setting correspond to `twisted geometries' \cite{newSpezialeFreidel}. Let us see how.

An element $H_{ab}$ of $SL(2,\mbb{C})$ can be written in terms of a positive real number $\eta_{ab}$ and two \emph{unrelated} $SU(2)$ group elements $g_{ab}$ and $g_{ba}$ as\footnote{In the following $\sigma_i$ are hermitian Pauli matrices.} \cite{Carmeli}
\begin{equation}
H_{ab}= g_{ab}\, e^{\,\eta\, \frac{\sigma_3}{2}}\, g_{ba}^{-1}\;.
\label{eq:H=gg}
\end{equation} 
In turn, a $SU(2)$ group element can be uniquely written in terms of an angle $\tilde{\phi}$ and a unit-vector $\vec{n}$. Let us define $\vec{n}$ via its inclination and azimuth
\begin{equation}
\vec{n}=\big(\sin \theta \cos \phi,\, \sin \theta \sin \phi,\, \cos \theta\big)\;,
\label{eq:vec n}
\end{equation}
and introduce the associated group element $n\in SU(2)$ defined as
\begin{equation}
n=e^{-i\phi \frac{\sigma_3}{2}}\, e^{-i\theta \frac{\sigma_2}{2}}\;.
\label{eq:n}
\end{equation}
Then the $SU(2)$ group element $g$ is given by $g= n\, e^{+i\tilde{\phi}\frac{\sigma_3}{2}}$. Using such parametrization in (\ref{eq:H=gg}) we finally find
\begin{equation}
H_{ab}=n_{ab}\, e^{-i z_{ab} \frac{\sigma_3}{2}}\, n_{ba}^{-1}\;.
\label{eq:H}
\end{equation}
with $z_{ab}=\xi_{ab}+i\eta_{ab}$ and $\xi_{ab}=\tilde{\phi}_{ba}-\tilde{\phi}_{ab}$. Therefore, for each link we have as labels the set $(\vec{n}_{ab},\vec{n}_{ba},\xi_{ab},\eta_{ab})$. These variables admit the following classical interpretation: a link connects two nodes living inside two adjacent chunks of space; the interface between them is a surface dual to the link; let us choose a frame in each of the two chunks; the variable $\vec{n}_{ab}$ can be interpreted as the (unit-)flux of the electric field $E^i$ in the chunk $a$ through the surface; similarly $\vec{n}_{ba}$ can be viewed as the flux in $b$ through this surface. In general, the two vectors are different as we have not chosen the same frame. There is a rotation $R$ such that $R\,\vec{n}_{ab}=-\vec{n}_{ba}$. The product $R\, e^{-i \xi \vec n_{ab}\cdot \frac{\vec{\sigma}}{2}}\in SU(2)$ can be understood as the holonomy of the Ashtekar-Barbero connection, $A^i=\Gamma^i+\gamma K^i$. Finally, the positive parameter $\eta_{ab}$ can be related to the area of the surface, i.e. to the spin $j_{ab}$.

\section{Semiclassical properties}
In order to test and strengthen our geometric interpretation of the $SL(2,\mbb{C})$ labels, in the following we study the asymptotics of coherent spin-networks for large parameter $\eta_{ab}$. This allows to test the proposal against candidate semiclassical states that have been studied previously.\\

The state (\ref{eq:proposal}) can be expanded on the spin-network basis $\Psi_{\Gamma,j_{ab},i_a}(h_{ab})$. Its components $f_{j_{ab},i_a}$,
\begin{equation}
\Psi_{\Gamma,H_{ab}}(h_{ab})=\sum_{j_{ab}}\sum_{i_a}\,f_{j_{ab},i_a}\;\Psi_{\Gamma,j_{ab},i_a}(h_{ab})
\label{eq:basis}
\end{equation} 
are given by\footnote{The notation $\cdot$ in (\ref{eq:components}) stands for a contraction of dual spaces. To be more explicit we recall that, if $V^{(j)}$ is the vector space where the representation $j$ of $SU(2)$ acts, then the tensor product of representations $D^{(j_e)}(h_e)$ lives in $\otimes_e (V^{(j_e)*}\otimes V^{(j_e)})$ while the tensor product of intertwiners lives precisely in the dual of this space. An orthonormal basis in intertwiner space is denoted $v_{i_a}$.}\footnote{Notice that here the sum over spins runs over half-integers \emph{including zero}. Thus, strictly speaking, a coherent spin-network does not live on a single graph $\Gamma$ but on a superposition of all the subgraphs of $\Gamma$.}
\begin{equation}
f_{j_{ab},i_a}=\Big(\prod_{ab}(2j_{ab}+1)e^{-j_{ab}(j_{ab}+1)t_{ab}} D^{(j_{ab})}(H_{ab})\Big)\cdot\Big(\prod_a v_{i_a}\Big)\;,
\label{eq:components}
\end{equation}
where from now on $H_{ab}$ is given by (\ref{eq:H}). Here we are interested in its asymptotics for $\eta_{ab}\gg 1$. First of all, notice that, in the limit $\eta_{ab}\to +\infty$, we have the following asymptotic behavior
\begin{equation}
D^{(j_{ab})}(e^{-i z_{ab} \frac{\sigma_3}{2}})^m_{\;\;m'}=\delta^m_{\;\;m'}e^{-i m z_{ab}}= \delta^m_{\;\;m'} \,e^{+ \eta_{ab} j_{ab}}\,\Big(\delta_{m,j_{ab}}e^{-i \xi_{ab} j_{ab}}+O(e^{-\eta_{ab}}) \Big)\;.
\end{equation}
Therefore, introducing the projector $P_+=|j_{ab},+j_{ab}\rangle\langle j_{ab},+j_{ab}|$ onto the highest magnetic number, we can write it as 
\begin{equation}
D^{(j_{ab})}(e^{-i z_{ab} \frac{\sigma_3}{2}})\approx e^{-i \xi_{ab} j_{ab}}e^{+  \eta_{ab} j_{ab}}P_+\;.
\end{equation}
Recall that the coherent intertwiners $\Phi_a(\vec{n}_{ab})$ introduced by Livine and Speziale \cite{Livine:2007vk} have components on a orthonormal basis $v_{i_a}$ in intertwiner space, $\Phi_a(\vec{n}_{ab})^{m_1\cdots m_r}=\sum_{i_a} \Phi_{i_a}(\vec{n}_{ab})\, v_{i_a}^{m_1\cdots m_r}$, given by
\begin{equation}
\Phi_{i_a}(\vec{n}_{ab})=v_{i_a}\cdot\Big({\bigotimes}_b |j_{ab},\vec{n}_{ab}\rangle\Big)
\label{eq:coherent intertwiners}
\end{equation}
where $|j_{ab},\vec{n}_{ab}\rangle=n_{ab}|j_{ab},+j_{ab}\rangle$. 
Moreover, notice that
\begin{equation}
-j(j+1)t+ j\,\eta=-\big(j-\frac{\eta-t}{2t}\big)^2\, t+\frac{(\eta-t)^2}{4t}\;.
\label{eq:quadratic}
\end{equation}
Therefore, up to an overall normalization of the state, we find the following asymptotics for our states:
\begin{equation}
f_{j_{ab},i_a}\approx \;\Big(\prod_{ab}(2j_{ab}+1) \exp(-\frac{(j_{ab}-j^0_{ab})^2}{2\sigma^0_{ab}})\,e^{-i \xi_{ab} j_{ab}} \Big)\;\Big(\prod_{a}\Phi_{i_a}(n_{ab}) \Big)\;
\label{eq:asymptotics}
\end{equation}
with 
\begin{equation}
(2j^0_{ab}+1)\equiv \frac{\eta_{ab}}{t_{ab}}\quad \text{and}\qquad \sigma^0_{ab}\equiv \frac{1}{2 t_{ab}}\;.
\label{eq:eta-t}
\end{equation}
Finally, introducing spin-networks with nodes labeled by coherent intertwiners as in \cite{Bianchi:2009ri},
\begin{equation}
\Psi_{\Gamma,j_{ab},\Phi_a(\vec{n}_{ab})}(h_{ab})=\sum_{i_a}\Big(\prod_a \Phi_{i_a}(\vec{n}_{ab})\Big)\;\Psi_{\Gamma,j_{ab},i_{a}}(h_{ab})\;,
\end{equation}
we find that the coherent spin-networks considered in this paper, for large $\eta_{ab}$, are given by the following superposition
\begin{equation}
\Psi_{\Gamma,H_{ab}}(h_{ab})\approx \sum_{j_{ab}}\Big(\prod_{ab}(2j_{ab}+1) \exp(-\frac{(j_{ab}-j^0_{ab})^2}{2\sigma^0_{ab}})\, e^{-i \xi_{ab} j_{ab}} \Big)\;\Psi_{\Gamma,j_{ab},\Phi_a(\vec{n}_{ab})}(h_{ab})\;.
\end{equation}
These are exactly the states considered by the authors as boundary semiclassical states in the analysis of the graviton propagator \cite{Bianchi:2009ri}. There, the graph $\Gamma$ is assumed to be the one dual to the boundary of a topological $4$-simplex, the quantities $j^0_{ab}$ and $\vec{n}_{ab}$ are areas and normals of faces of tetrahedra chosen so to reproduce the intrinsic geometry of the boundary of a \emph{regular Euclidean $4$-simplex}. Moreover, the parameters $\xi_{ab}$ are chosen so to reproduce its extrinsic curvature. The analysis of the correlation function of metric operators confirms that the appropriate value is $\xi_{ab}=\gamma K_{ab}=\gamma \arccos(-1/4)$. This result confirms the geometric interpretation of our variables and extends the validity of the semiclassical states used in \cite{Bianchi:2009ri} well beyond the simplicial setting: coherent spin-networks are defined in full LQG.\\

In order to better test the interpretation of our variables, we consider a rather simple example: the coherent loop. This example allows us to discuss the importance of the appropriate choice of heat-kernel time $t_{ab}$ in (\ref{eq:proposal}).

When the graph is given by a loop $\gamma$, the dependence of the state on the normals $\vec{n}$ in (\ref{eq:proposal}) drops out and the state is simply labeled by a complex number $z=\xi+i\eta$,
\begin{equation}
\Psi_{\gamma,z}(h)=\sum_j e^{-j(j+1)t}\, \frac{\sin ((2j+1) z/2)}{\sin(z/2)}\, \chi^{(j)}(h)\;.
\label{eq:psi z}
\end{equation}
For large $\eta$, we find
\begin{equation}
\Psi_{\gamma,\xi+i\eta}(h)=\sum_j\exp\big(-\frac{(j-j_0)^2}{2\sigma_0}\big)e^{-i \xi j} \chi^{(j)}(h)\;
\label{eq:loop}
\end{equation}
with $j_0$ and $\sigma_0$ given in terms of $\eta$ and $t$ by (\ref{eq:eta-t}). Now we compute the expectation value of the area operator $\mathscr{A}$ for a surface that is punctured once by the loop. As well known, we have
\begin{equation}
\hat{\mathscr{A}}\,\chi^{(j)}(h)=\gamma L_P^2\,\sqrt{j(j+1)}\;\chi^{(j)}(h)\;.
\label{eq:area operator}
\end{equation}
Therefore
\begin{equation}
\hat{\mathscr{A}}\,\Psi_{\gamma,\xi+i\eta}(h)=\gamma L_P^2\,\sum_j\exp\big(-\frac{(j-j_0)^2}{2\sigma_0}\big)\,e^{-i \xi j}\;\sqrt{j(j+1)}\; \chi^{(j)}(h)\;.
\label{eq:A Psi}
\end{equation}
In the limit of large $\eta$ \emph{and} large $j_0$, the expectation value of the area operator is easily computed
\begin{equation}
\langle\mathscr{A}\rangle=\frac{(\Psi_{\gamma,\xi+i\eta},\,\hat{\mathscr{A}}\,\Psi_{\gamma,\xi+i\eta})}{(\Psi_{\gamma,\xi+i\eta},\Psi_{\gamma,\xi+i\eta})}=\gamma L_P^2\,\sqrt{j_0(j_0+1)}\;
\end{equation}
and confirms the interpretation of $\eta$ as the quantity that prescribes the expectation value of the area. 

Now we consider the other observable acting on the Hilbert space $\mc{K}_\gamma$: the Wilson loop operator $W_\gamma$. Recall that it acts on basis vectors as
\begin{equation}
\hat{W}_\gamma \,\chi^{(j)}(h)\equiv \chi^{(\frac{1}{2})}(h)\,\chi^{(j)}(h)=\chi^{(j+\frac{1}{2})}(h)+\chi^{(j-\frac{1}{2})}(h)\;.
\label{eq:W chi}
\end{equation}
As a result, we find
\begin{equation}
\langle W_\gamma \rangle= 2\, \cos(\xi/2)\;e^{-\frac{t}{8}}.
\end{equation}
Therefore, in the limit $t\to 0$ compatible with $\eta$ and $j_0$ large, the parameter $\xi$ identifies the conjugacy class of the group element $h_0$ where the Ashtekar-Barbero connection is peaked on. According to the Aharonov-Bohm picture of LQG \cite{Bianchi:2009tj}, the angle $\xi$ is thus the expectation value of the flux of the magnetic field through a line defect encircled by the loop $\gamma$.

Similarly, we can compute the dispersions of the area operator and of the Wilson loop. We find
\begin{equation}
\Delta\mathscr{A}\equiv\sqrt{\langle\mathscr{A}^2\rangle-\langle\mathscr{A}\rangle^2}=\frac{1}{2}\gamma L_P^2\;\sqrt{2\sigma_0}\;,
\end{equation}
and
\begin{equation}
\Delta W_\gamma \equiv\sqrt{\langle W_\gamma^2\rangle-\langle W_\gamma\rangle^2}=\sin (\xi/2)\;\frac{1}{\sqrt{2\sigma_0}}\;.
\end{equation}
Now notice that, due to the relation (\ref{eq:eta-t}), the limit ``large $\eta$ and large $j_0$'' can be attained only if we assume that $t$ scales with $j_0$ as
\begin{equation}
t\sim (j_0)^k \qquad \text{with}\quad k>-1\;.
\label{eq:t-j0}
\end{equation}
Moreover, as the area and the Wilson loop are non-commuting operators, we cannot make both their dispersions vanish at the same time. Small heat-kernel time means that the state is sharply peaked on the holonomy, while large heat-kernel time means that the state is sharply peaked on the spin. A good requirement of semiclassicality is that the relative dispersions of both operators vanish in the large $j_0$ limit. Using the results derived above, we find the following behavior for relative dispersions:
\begin{equation}
\frac{\Delta\mathscr{A}}{\langle\mathscr{A}\rangle}\sim (j_0)^{-\frac{k+2}{2}}\;\qquad \text{and}\qquad \frac{\Delta W_\gamma}{\langle W_\gamma\rangle}\sim (j_0)^{\frac{k}{2}}\;.
\label{eq:dA/A}
\end{equation}
The first requires $k>-2$ and the second $k<0$. Taking into account the three bounds (\ref{eq:t-j0})-(\ref{eq:dA/A}) we find that the coherent loop behaves semiclassically when the heat-kernel time scales as $(j_0)^k$ with $-1<k<0$. For instance, the choice $t=1/\sqrt{j_0}$ guarantees the semiclassicality of the state.

\section{Resolution of the identity}
In the previous section we focused on the properties of semiclassicality of coherent spin-networks: peakedness on a classical configuration with small dispersions. In this section we discuss their \emph{coherence} properties: for a given choice of parameters $t_{ab}$, coherent spin-networks provide a holomorphic representation for Loop Quantum Gravity. This result was obtained long ago by Ashtekar, Lewandowski, Marolf, Mour\~ao and Thiemann \cite{Ashtekar:1994nx} and is based on the Segal-Bargmann transform for compact Lie groups introduced by Hall \cite{Hall1994}. Here we report their result in the formalism of this paper and comment on its relevance for the analysis of the semiclassical behavior of Loop Quantum Gravity.\\

Let us consider the $SL(2,\mbb{C})$ heat-kernel\footnote{The $SL(2,\mbb{C})$ heat-kernel $F_t(H,H_0)$ is not to be confused with the analytic continuation to $SL(2,\mbb{C})$ of the $SU(2)$ heat-kernel $K_t(h,h_0)$.} $F_t(H)$ and introduce a function $\Omega_t(H)$ on $SL(2,\mbb{C})$ given by
\begin{equation}
\Omega_t(H)=\int_{SU(2)} F_t( H g)\, dg\;.
\label{eq:omega t}
\end{equation}
This function is just the heat-kernel on $SL(2,\mbb{C})/SU(2)$, regarded as a $SU(2)$-invariant function on $SL(2,\mbb{C})$. A key result of Hall \cite{Hall1994} 
is that the delta function on $SU(2)$ can be written in terms of the following $SL(2,\mbb{C})$ integral
\begin{equation}
\delta(h,h')=\int_{SL(2,\mbb{C})} K_t(h H^{-1})\; \overline{K_t(h' H^{-1})}\;\Omega_{2t}(H)\;dH\;,
\label{eq:delta link}
\end{equation}
where $dH$ is the Haar measure on $SL(2,\mbb{C})$. This expression admits a straightforward generalization in terms of coherent spin-networks.

We recall that, in the holonomy representation, the identity operator ${\bf 1}_\Gamma$ on the Hilbert space $\mc{K}_\Gamma$ is given by the distribution $\delta_\Gamma(h_{ab},h'_{ab})$ on $SU(2)^L/SU(2)^N$. It can be written in terms of the spin-network basis as
\begin{align}
\delta_\Gamma(h_{ab},h'_{ab})=&\;\int(\prod_a dg_a)(\prod_a dg'_a)\;\prod_{ab}\delta(g_a^{-1} h_{ab} g_b, {g'}_a^{-1} h'_{ab} g'_b) \label{eq:delta gamma}\\
=&\;\sum_{j_{ab} i_a}\Psi_{\Gamma, j_{ab}, i_a}(h_{ab})\; \overline{\Psi_{\Gamma, j_{ab}, i_a}(h'_{ab})}\;.\nonumber
\end{align}
The resolution of the identity for coherent spin-networks is given by
\begin{equation}
\delta_\Gamma(h_{ab},h'_{ab})=\int_{SL(2,\mbb{C})^L}\Psi_{\Gamma, H_{ab}}(h_{ab})\; \overline{\Psi_{\Gamma, H_{ab}}(h'_{ab})}\;\;\; \big(\prod_{ab} \Omega_{2t_{ab}}(H_{ab})\, dH_{ab}\big)
\label{eq:resolution}
\end{equation}
with the measure on $SL(2,\mbb{C})^L$ that factors in a product of measures per link given by the $SU(2)$-averaged heat-kernel for $SL(2,\mbb{C})$ at time $2t$, times the Haar measure $dH_{ab}$. Expression (\ref{eq:resolution}) for the resolution of the identity can be easily proved using formula (\ref{eq:delta link}), the definition of coherent spin-networks (\ref{eq:proposal}) and expression (\ref{eq:delta gamma}) for $\delta_\Gamma(h_{ab},h'_{ab})$.\\

As shown in \cite{Ashtekar:1994nx}, coherent spin-networks provide a Segal-Bargmann transform for Loop Quantum Gravity. 
In fact, given a state $\Psi_{\Gamma,f}(h_{ab})$, its scalar product with a coherent spin-network $\Psi_{\Gamma,H_{ab}}(h_{ab})$ defines a function $\Phi_{\Gamma,f}(H_{ab})$ that is holomorphic in $H_{ab}$,
\begin{equation}
\Phi_{\Gamma,f}(H_{ab})=\int_{SU(2)^L}\Psi_{\Gamma,H_{ab}}(h_{ab})\;\overline{\Psi_{\Gamma,f}(h_{ab})}\;\prod_{ab}dh_{ab}\;,
\label{eq:trasform}
\end{equation}
and belongs to the Hilbert space $\mc{H}L^2(SL(2,\mbb{C})^L,(\Omega_{2t} dH)^L)$ of holomorphic functions normalizable with respect to the measure $(\Omega_{2t} dH)^L$. Moreover, from expression (\ref{eq:resolution}) follows that the trasform preserves the scalar product,
\begin{equation}
\int_{SU(2)^L}\hspace{-1.5em}\Psi_{\Gamma,f_1}(h_{ab})\;\overline{\Psi_{\Gamma,f_2}(h_{ab})}\;\prod_{ab}dh_{ab}\;\;\;=\;\int_{SL(2,\mbb{C})^L}\hspace{-1.5em}\Phi_{\Gamma,f_1}(H_{ab})\;\overline{\Phi_{\Gamma,f_2}(H_{ab})}\;\;\big(\prod_{ab} \Omega_{2t_{ab}}(H_{ab})\, dH_{ab}\big)\;.\nonumber
\end{equation}
What is now available is a representation for Loop Quantum Gravity where states are functions of classical variables $H_{ab}$ that admit a clear geometric interpretation in terms of areas, extrinsic angles and normals, $(\eta_{ab},\xi_{ab},\vec{n}_{ab})$, the variables generally used in the Spin Foam setting.

\section{Conclusions}
We have discussed a proposal of coherent states for Loop Quantum Gravity and shown that, in a specific limit, they reproduce the states used in the Spin Foam framework. Moreover, these states coincide with Thiemann's complexifier coherent states with the natural choice of complexifier operator, a rather specific choice of heat-kernel time and a clear geometrical interpretation for their $SL(2,\mbb{C})$ labels. 

Coherent spin-networks are candidate semiclassical states for full Loop Quantum Gravity. Given a space-time metric (for instance the Minkowski one), we can identify an intrinsic and extrinsic metric on a spatial slice $\Sigma$. Then, we can consider a cellular decomposition of $\Sigma$ and a graph $\Gamma$ embedded in $\Sigma$ and dual to the decomposition. The data captured by the graph is easy to determine: we can smear the Ashtekar-Barbero connection on links of the graph and the electric field on surfaces dual to links. This procedure determines a finite amount of data that can be used as labels for the coherent state. In the case of a simplicial decomposition, we know that this data correspond to a Regge geometry with dislocations. These geometries are studied in \cite{newSpezialeFreidel} and are called `twisted'.

The large area asymptotics of coherent spin-networks reproduces Gaussian superpositions on spins, times Livine-Speziale coherent intertwiners at nodes. A Gaussian on spins in the asymptotics was to be expected and is not surprising as the motivation of Hall-type coherent states is exactly to generalize Gaussians to the group manifold. On the other hand, the recovery of Livine-Speziale coherent intertwiners at nodes is a surprising feature of coherent spin-networks. This property was not suspected before and is the central result of this paper.

The fact that, in the large spin limit, the states we consider reproduce the Livine-Speziale coherent intertwiners on nodes guaranties that they are actually peaked on a classical expectation value of non-commuting geometric operators. For instance, we know that the expectation value of the volume of a region containing a $4$-valent node is given by the classical volume of a Euclidean tetrahedron with the normals to faces and the areas as prescribed by the labels of the coherent spin-network. What needs to be better understood is the relation and the origin of the tension with the results of Flori and Thiemann \cite{Flori:2008nw,Flori:2009rw} where they claim that only nodes of valency $6$ can have a semiclassical behavior.\\

There is a number of possible developments that we can envision at this early stage. 
An aspect that needs further investigation regards the redundancy of the labels of these states.  The situation is analogous to the one of the labels of the Livine-Speziale coherent intertwiners (four unit-normals) \cite{Livine:2007vk} as opposed to the constrained labels of the same states obtained by Conrady and Freidel via geometric quantization of a classical tetrahedron \cite{Conrady:2009px,Freidel:2009nu}. Similarly, it is possible that the coherent states obtained via geometric quantization of the phase space of LQG associated to a graph actually coincide with a subset of the coherent states introduced here via heat-kernel methods. This would be an instance of Guillemin-Sternberg's `quantization commuting with reduction'.
Notice that the situation here is slightly more involved: geometric quantization seemingly knows nothing about heat kernels. Nevertheless, explicit computations by Hall \cite{Hall2002} show that amazingly the two approaches give precisely the same coherent states in all the studied cases. This aspect deserves to be understood better.

A surprising property of the states we have discussed is that they bring together so many (apparently conflicting) ideas that have been proposed in the search for semiclassical states in Loop Quantum Gravity. We consider this convergence to be a measure of the robustness of the theory. 

\section*{Acknowledgments}
We thank Carlo Rovelli for helpful discussions. E.M. gratefully acknowledges support from Fondazione A. Della Riccia. The work of E.B. is supported by a Marie Curie Intra-European Fellowship within the 7th European Community Framework Programme.

\appendix

\section{Proof of formula (\ref{eq:delta link})}\label{app:proof}
The representation of the delta function of $SU(2)$ in terms of an integral on $SL(2,\mbb{C})$, formula (\ref{eq:delta link}), is a key ingredient in the proof of the resolution of the identity provided by coherent spin-networks. To make the paper self-contained, in this appendix we report an elementary proof of formula (\ref{eq:delta link}). The proof is by direct computation and is similar to the derivation of \cite{Thiemann:2000ca} (section 4.4). A more general proof for compact Lie groups can be found in \cite{Hall1994} (section 7).\\

Let us use the polar decomposition
\begin{equation}
H=\,g \,e^{\,\vec{p}\,\cdot \frac{\vec{\sigma}}{2}}
\label{eq:polar}
\end{equation}
to parametrize an element $H$ of $SL(2,\mbb{C})$ in terms of an element $g$ of $SU(2)$ and a vector $\vec{p}$ in $\mbb{R}^3$. In these variables, the Haar measure $dH$ on $SL(2,\mbb{C})$ factors into a $SU(2)$ term and a $\mbb{R}^3$ term
\begin{equation}
dH\;=\; \frac{(\sinh |\vec{p}\,|)^2}{|\vec{p}\,|^2} d^3\vec{p}\; dg\;,
\label{eq:dH}
\end{equation}
where $dg$ is the Haar measure on $SU(2)$ and $d^3\vec{p}$ is the Lebesgue measure on $\mbb{R}^3$.

The $SU(2)$-averaged heat kernel on $SL(2,\mbb{C})$ coincides with the heat kernel on the hyperboloid $H^3=SL(2,\mbb{C})/SU(2)$. Its explicit form in terms of the variables (\ref{eq:polar}) can be found in \cite{Gang1968} and is given by
\begin{equation}
\Omega_t(g \,e^{\,\vec{p}\,\cdot \frac{\vec{\sigma}}{2}})=\frac{1}{(\pi t)^{3/2}}e^{-t/4}\; \frac{|\vec{p}\,|}{\sinh |\vec{p}\,|} e^{-|\vec{p}\,|^2/t}\;.
\label{eq:gangolli}
\end{equation}
Therefore, the measure in the resolution of the identity (\ref{eq:delta link}) is given by
\begin{equation}
\Omega_{2t}(H_{ab})\, dH_{ab}=\rho_t(|\vec{p}\,|)\;d^3\vec{p}\;dg
\label{eq:O2t}
\end{equation}
where
\begin{equation}
\rho_t(|\vec{p}\,|)=\frac{1}{(2\pi t)^{3/2}}e^{-t/2}\;\frac{\sinh |\vec{p}\,|}{|\vec{p}\,|}\;e^{-\frac{|\vec{p}\,|^2}{2t}}\;.
\label{eq:rho}
\end{equation}

Now we want to compute the integral that appears on the rhs of (\ref{eq:delta link}). Using the Peter-Weyl expansion of the heat-kernel we find
\begin{equation}
\int_{SL(2,\mbb{C})}\hspace{-1.5em} K_t(h H^{-1})\; \overline{K_t(h' H^{-1})}\;\Omega_{2t}(H)\;dH\;= \sum_{j, j'} (2j+1)(2j'+1) e^{-j(j+1)t}e^{-j'(j'+1)t}\; f_{j,j'}(h,h')\;, 
\label{eq:KK}
\end{equation}
where the coefficients in the sum are given by
\begin{equation}
f_{j,j'}(h,h')=\int_{SL(2,\mbb{C})} \chi^{(j)}(h H^{-1})\; \overline{\chi^{(j')}(h' H^{-1})}\;\Omega_{2t}(H)\;dH\;.
\label{eq:fjj}
\end{equation}
This quantity can be computed in two steps: (i) first we integrate over the subgroup $SU(2)$ and obtain
\begin{equation}
f_{j,j'}(h,h')=\frac{\delta^{j,j'}}{2j+1}\,\Tr(D^{j}(h {h'}^{-1}) A)
\label{eq:fjj1}
\end{equation}
where $A$  is a $(2j+1)\times(2j+1)$ matrix. Then, (ii) we compute the matrix $A$. It is given by the following integral over $\mbb{R}^3$
\begin{equation}
A=\int_{\mbb{R}^3}D^{(j)}(e^{-\vec{p}\cdot\frac{\vec{\sigma}}{2}})\;\rho_t(|\vec{p}\,|)\,d^3\vec{p}\;.
\label{eq:A}
\end{equation}
Notice that the matrix $A$ commutes with the irreducible representation $j$ of $SU(2)$. Therefore it has to be a multiple of the identity
\begin{equation}
A=\frac{c}{2j+1}\mbf{1}
\label{eq:id}
\end{equation}
with the constant $c$ given by the trace of the matrix. Such constant can be computed explicitly performing the integral and is given by
\begin{equation}
c=\Tr A=4\pi\int_0^\infty\frac{\sinh((2j+1)p )}{\sinh(p)}\;\rho_t(p)\,dp\;=\;(2j+1)\;e^{+j(j+1)2t}\;.
\label{eq:c}
\end{equation}
Therefore the integral (\ref{eq:fjj}) is simply given by
\begin{equation}
f_{j,j'}(h,h')=\;\delta^{j,j'}\;e^{+j(j+1)2t}\;\chi^{(j)}(h {h'}^{-1})\;.
\label{eq:fjj2}
\end{equation}
Inserting this result into the rhs of expression (\ref{eq:KK}) we find the Peter-Weyl expansion of the delta-function,
\begin{equation}
\delta(h,h')=\sum_j (2j+1) \chi^{(j)}(h {h'}^{-1})\;.
\label{eq:dhh}
\end{equation} 
This proves expression (\ref{eq:delta link}).


\begin{thebibliography}{10}

\bibitem{LQG}
C.~Rovelli, {\em Quantum Gravity}.
\newblock Cambridge University Press, 2004.

\bibitem{Smolin:2004sx}
L.~Smolin, ``{An invitation to loop quantum gravity},''
\href{http://arXiv.org/abs/hep-th/0408048}{{\tt hep-th/0408048}}.

\bibitem{Ashtekar:2004eh}
A.~Ashtekar and J.~Lewandowski, ``{Background independent quantum gravity: A
  status report},'' {\em Class. Quant. Grav.} {\bf 21} (2004) R53,
\href{http://arXiv.org/abs/gr-qc/0404018}{{\tt gr-qc/0404018}}.

\bibitem{Thiemann:2007zz}
T.~Thiemann, ``{Modern canonical quantum general relativity},''. Cambridge, UK:
  Cambridge Univ. Pr. (2007) 819 p.

\bibitem{Engle:2007wy}
J.~Engle, E.~Livine, R.~Pereira, and C.~Rovelli, ``{LQG vertex with finite
  Immirzi parameter},'' {\em Nucl. Phys.} {\bf B799} (2008) 136--149,
\href{http://arXiv.org/abs/0711.0146}{{\tt 0711.0146}}.

\bibitem{Freidel:2007py}
L.~Freidel and K.~Krasnov, ``{A New Spin Foam Model for 4d Gravity},'' {\em
  Class. Quant. Grav.} {\bf 25} (2008) 125018,
\href{http://arXiv.org/abs/0708.1595}{{\tt 0708.1595}}.

\bibitem{Kaminski:2009fm}
W.~Kaminski, M.~Kisielowski, and J.~Lewandowski, ``{Spin-Foams for All Loop
  Quantum Gravity},''  {\em Class.\ Quant.\ Grav.\ }  {\bf 27} (2010) 095006.
\href{http://arXiv.org/abs/0909.0939}{{\tt 0909.0939}}.

\bibitem{Rovelli:2005yj}
C.~Rovelli, ``{Graviton propagator from background-independent quantum
  gravity},'' {\em Phys. Rev. Lett.} {\bf 97} (2006) 151301,
\href{http://arXiv.org/abs/gr-qc/0508124}{{\tt gr-qc/0508124}}.

\bibitem{Bianchi:2006uf}
E.~Bianchi, L.~Modesto, C.~Rovelli, and S.~Speziale, ``{Graviton propagator in
  loop quantum gravity},'' {\em Class. Quant. Grav.} {\bf 23} (2006)
  6989--7028,
\href{http://arXiv.org/abs/gr-qc/0604044}{{\tt gr-qc/0604044}}.

\bibitem{Livine:2006it}
E.~R. Livine and S.~Speziale, ``{Group Integral Techniques for the Spinfoam
  Graviton Propagator},'' {\em JHEP} {\bf 11} (2006) 092,
\href{http://arXiv.org/abs/gr-qc/0608131}{{\tt gr-qc/0608131}}.

\bibitem{Alesci:2007tx}
E.~Alesci and C.~Rovelli, ``{The complete LQG propagator: I. Difficulties with
  the Barrett-Crane vertex},'' {\em Phys. Rev.} {\bf D76} (2007) 104012,
\href{http://arXiv.org/abs/0708.0883}{{\tt 0708.0883}}.

\bibitem{Alesci:2007tg}
E.~Alesci and C.~Rovelli, ``{The complete LQG propagator: II. Asymptotic
  behavior of the vertex},'' {\em Phys. Rev.} {\bf D77} (2008) 044024,
\href{http://arXiv.org/abs/0711.1284}{{\tt 0711.1284}}.

\bibitem{Alesci:2008ff}
E.~Alesci, E.~Bianchi, and C.~Rovelli, ``{LQG propagator: III. The new
  vertex},'' {\em Class. Quant. Grav.} {\bf 26} (2009) 215001,
\href{http://arXiv.org/abs/0812.5018}{{\tt 0812.5018}}.

\bibitem{Bianchi:2009ri}
E.~Bianchi, E.~Magliaro, and C.~Perini, ``{LQG propagator from the new spin
  foams},'' {\em Nucl. Phys.} {\bf B822} (2009) 245--269,
\href{http://arXiv.org/abs/0905.4082}{{\tt 0905.4082}}.

\bibitem{Livine:2007vk}
E.~R. Livine and S.~Speziale, ``{A new spinfoam vertex for quantum gravity},''
  {\em Phys. Rev.} {\bf D76} (2007) 084028,
\href{http://arXiv.org/abs/0705.0674}{{\tt 0705.0674}}.

\bibitem{Conrady:2009px}
F.~Conrady and L.~Freidel, ``{Quantum geometry from phase space reduction},''
{\em J.\ Math.\ Phys.\ } {\bf 50}, 123510 (2009),
\href{http://arXiv.org/abs/0902.0351}{{\tt 0902.0351}}.

\bibitem{Freidel:2009nu}
L.~Freidel, K.~Krasnov, and E.~R. Livine, ``{Holomorphic Factorization for a
  Quantum Tetrahedron},''
 {\em Commun.\ Math.\ Phys.\ } {\bf 297}, 45 (2010)
\href{http://arXiv.org/abs/0905.3627}{{\tt 0905.3627}}.

\bibitem{SpezialeBeijing}
S.~Speziale, ``{Twisted geometries: a geometric parametrization of SU(2) phase
  space},'' {\em talk given at Loops 2009 in Beijing}.

\bibitem{newSpezialeFreidel}
  L.~Freidel and S.~Speziale,
  ``Twisted geometries: A geometric parametrisation of SU(2) phase space,''
\href{http://arXiv.org/abs/1001.2748}{{\tt 1001.2748}}.

\bibitem{Thiemann:2000bw}
T.~Thiemann, ``{Gauge field theory coherent states (GCS). I: General
  properties},'' {\em Class. Quant. Grav.} {\bf 18} (2001) 2025--2064,
\href{http://arXiv.org/abs/hep-th/0005233}{{\tt hep-th/0005233}}.

\bibitem{Thiemann:2000ca}
T.~Thiemann and O.~Winkler, ``{Gauge field theory coherent states (GCS). II:
  Peakedness properties},'' {\em Class. Quant. Grav.} {\bf 18} (2001)
  2561--2636,
\href{http://arXiv.org/abs/hep-th/0005237}{{\tt hep-th/0005237}}.

\bibitem{Thiemann:2000bx}
T.~Thiemann and O.~Winkler, ``{Gauge field theory coherent states (GCS) III:
  Ehrenfest theorems},'' {\em Class. Quant. Grav.} {\bf 18} (2001) 4629--4682,
\href{http://arXiv.org/abs/hep-th/0005234}{{\tt hep-th/0005234}}.

\bibitem{Thiemann:2000by}
T.~Thiemann and O.~Winkler, ``{Gauge field theory coherent states (GCS). IV:
  Infinite tensor product and thermodynamical limit},'' {\em Class. Quant.
  Grav.} {\bf 18} (2001) 4997--5054,
\href{http://arXiv.org/abs/hep-th/0005235}{{\tt hep-th/0005235}}.

\bibitem{Sahlmann:2001nv}
H.~Sahlmann, T.~Thiemann, and O.~Winkler, ``{Coherent states for canonical
  quantum general relativity and the infinite tensor product extension},'' {\em
  Nucl. Phys.} {\bf B606} (2001) 401--440,
\href{http://arXiv.org/abs/gr-qc/0102038}{{\tt gr-qc/0102038}}.

\bibitem{Thiemann:2002vj}
T.~Thiemann, ``{Complexifier coherent states for quantum general relativity},''
  {\em Class. Quant. Grav.} {\bf 23} (2006) 2063--2118,
\href{http://arXiv.org/abs/gr-qc/0206037}{{\tt gr-qc/0206037}}.

\bibitem{Bahr:2007xa}
B.~Bahr and T.~Thiemann, ``{Gauge-invariant coherent states for Loop Quantum
  Gravity I: Abelian gauge groups},'' {\em Class. Quant. Grav.} {\bf 26} (2009)
  045011,
\href{http://arXiv.org/abs/0709.4619}{{\tt 0709.4619}}.

\bibitem{Bahr:2007xn}
B.~Bahr and T.~Thiemann, ``{Gauge-invariant coherent states for Loop Quantum
  Gravity II: Non-abelian gauge groups},'' {\em Class. Quant. Grav.} {\bf 26}
  (2009) 045012,
\href{http://arXiv.org/abs/0709.4636}{{\tt 0709.4636}}.

\bibitem{Flori:2008nw}
C.~Flori and T.~Thiemann, ``{Semiclassical analysis of the Loop Quantum Gravity
  volume operator: I. Flux Coherent States},''
\href{http://arXiv.org/abs/0812.1537}{{\tt 0812.1537}}.

\bibitem{Flori:2009rw}
C.~Flori, ``{Semiclassical analysis of the Loop Quantum Gravity volume
  operator: Area Coherent States},''
\href{http://arXiv.org/abs/0904.1303}{{\tt 0904.1303}}.

\bibitem{Hall1994}
B.~C. Hall, ``{The Segal-Bargmann Coherent State Transform for Compact Lie
  Groups},'' {\em J. Funct. Anal.} {\bf 122} (1994) 103--151.

\bibitem{Ashtekar:1994nx}
A.~Ashtekar, J.~Lewandowski, D.~Marolf, J.~Mour\~ao, and T.~Thiemann, ``{Coherent
  state transforms for spaces of connections},'' {\em J. Funct. Anal.} {\bf
  135} (1996) 519--551,
\href{http://arXiv.org/abs/gr-qc/9412014}{{\tt gr-qc/9412014}}.

\bibitem{Carmeli}
M.~Carmeli, {Group Theory and General Relativity: Representations of the
  Lorentz Group and Their Applications to the Gravitational Field}.
\newblock Imperial College Press, 2000.

\bibitem{Bianchi:2009tj}
E.~Bianchi, ``{Loop Quantum Gravity \`a la Aharonov-Bohm},''
\href{http://arXiv.org/abs/0907.4388}{{\tt 0907.4388}}.

\bibitem{Hall2002}
B.~C. Hall, ``{Geometric quantization and the generalized Segal-Bargmann
  transform for Lie groups of compact type},'' {\em Comm. Math. Phys.} {\bf
  226} (2002) 233--268.
  
\bibitem{Gang1968} R.~Gangolli, ``{Asymptotic behavior of spectra of compact quotients of certain symmetric spaces},'' {\em Acta Math.} {\bf 121} (1968), 151–-192.

\end{thebibliography}

\providecommand{\href}[2]{#2}\begingroup\raggedright\endgroup

\end{document}